\documentclass{aa}

\usepackage[caption=false]{subfig}
\usepackage[colorlinks,linkcolor=blue,citecolor=blue]{hyperref}
\usepackage[varg]{txfonts}
\usepackage{graphicx}

\DeclareMathAlphabet{\mathcal}{OMS}{cmsy}{m}{n}
\DeclareSymbolFont{largesymbols}{OMX}{cmex}{m}{n}

\newcommand{\Eiso}{E_{\rm{iso}}}
\newcommand{\Ep}{E_{\rm{p}}}
\newcommand{\Lp}{L_{\rm{p}}}
\newcommand{\obs}{\theta_{\rm{obs}}}
\newcommand{\jet}{\theta_{\rm{jet}}}
\newcommand{\ginj}{\gamma^{\prime}_{\rm{inj}}}

\begin{document}

\title{An analytic derivation of the empirical correlations of gamma-ray bursts}

\author{Fan Xu\inst{1},
    Yong-Feng Huang\inst{1,2}\thanks{e-mail: {\tt hyf@nju.edu.cn}},
    Jin-Jun Geng\inst{3},
    Xue-Feng Wu\inst{3},
    Xiu-Juan Li\inst{4} 
    and Zhi-Bin Zhang \inst{5}
}

\institute{
    \inst{1}{School of Astronomy and Space Science, Nanjing University, Nanjing 210023, People's Republic of China}\\
    \inst{2}{Key Laboratory of Modern Astronomy and Astrophysics (Nanjing University), Ministry of Education, People's Republic of China}\\
    \inst{3}{Purple Mountain Observatory, Chinese Academy of Sciences, Nanjing 210023, People's Republic of China}\\
    \inst{4}{School of Cyber Science and Engineering, Qufu Normal University, Qufu 273165, People's Republic of China}\\
    \inst{5}{College of Physics and Engineering, Qufu Normal University, Qufu 273165, People's Republic of China}\\
}

\date{}

\abstract {Empirical correlations between various key parameters
have been extensively explored ever since the discovery of
gamma-ray bursts (GRBs) and have been widely used as standard
candles to probe the Universe. The Amati relation and the Yonetoku
relation are two good examples that enjoyed special
attention. The former reflects the connection between the peak
photon energy ($\Ep$) and the isotropic $\gamma$-ray energy
release ($\Eiso$), while the latter links $\Ep$ with the isotropic
peak luminosity ($\Lp$), both in the form of a power-law function.
Most GRBs are found to follow these correlations well, but a
theoretical interpretation is still lacking. Some obvious outliers may be off-axis GRBs and may
follow correlations that are different from those at the on-axis.
Here we present a simple analytical derivation for the Amati
relation and the Yonetoku relation in the framework of the
standard fireball model, the correctness of which is then
confirmed by numerical simulations. The off-axis Amati relation
and Yonetoku relation are also derived. They differ markedly from the
corresponding on-axis relation. Our results reveal the
intrinsic physics behind the radiation processes of GRBs,
and they highlight the importance of the viewing angle in the empirical
correlations of GRBs.}


\titlerunning{Empirical correlations of GRBs}
\authorrunning{Xu et al.}
\maketitle

\section{Introduction} \label{sec:intro}
After about five decades of research, some
insights into gamma-ray bursts (GRBs) have been obtained. Generally speaking, there
are two different phases in GRBs: the prompt emission, and the
afterglow. The afterglow can be relatively well interpreted by the
external shock model
\citep{Meszaros..1997,Sari..1998,Huang..2006,Geng..2016,Lazzati..2018,Xu..2022}.
However, the radiation process of the prompt emission is still
under debate. Different models have been proposed to explain the
complicated prompt emission, such as the internal shock model
\citep{Rees..1994,Kobayashi..1997,Bosnjak..2009}, the dissipative
photosphere model \citep{Rees..2005,Peer..2011}, and the internal-collision-induced magnetic reconnection and turbulence (ICMART) model \citep{Zhang..2011,Zhang..2014}. The most commonly discussed
model is the internal shock model, which is naturally expected for
a highly variable central engine. In this model, the collision and
merger of shells create relativistic shocks to accelerate
particles. Then the accelerated particles will cause the
observed GRB prompt emission.

The spectrum of GRB prompt emission has traditionally been
described by the empirical Band function \citep{Band..1993}. It has
been suggested that synchrotron radiation may be responsible for
the non-thermal component \citep{Rees..1994,Sari..1998}. However,
for a standard fast-cooling spectrum, the theoretical low-energy
spectral index is too soft \citep{Sari..1998}. Several studies
have shown that this problem can be mitigated when a decaying magnetic
field and the detailed cooling process are considered
\citep{Uhm..2014,Zhang..2016,Geng..2018}. Some authors
argued that the low-energy index of the empirical Band function may
be misleading \citep{Burgess..2020}. They proposed that the
synchrotron emission mechanism can interpret most spectra and can
satisfactorily fit the observational data \citep{Zhang..2020}.

In addition to the spectrum, the diversity of GRB energy is another
mystery. For a typical long GRB (duration longer than $2$ s), the
isotropic energy is around $10^{52}-10^{54}$ erg. However, there
also exist some low-luminosity GRBs (LLGRBs) with energies down to
$10^{48}$ erg (GRB 980425) for long GRBs and $10^{46}$ erg (GRB
170817A) for short GRBs (duration shorter than $2$ s). Some
authors claimed that LLGRBs may form a distinct population of GRBs
\citep{Liang..2007}. However, a later study reported no clear
separation between LLGRBs and standard high-luminosity GRBs in a
larger GRB sample \citep{Sun..2015}. On the other hand, LLGRBs can naturally be explained by off-axis jets. Due to the relativistic
beaming effect, a GRB will become dimmer when the viewing angle
($\obs$) is larger than the jet opening angle ($\jet$)
\citep{Granot..2002,Huang..2002,Yamazaki..2003a}. The interesting
short GRB 170817A clearly proves that at least some LLGRBs are
viewed off-axis \citep{Granot..2017}.

The GRB empirical relations can help us understand the physical
nature of GRBs. The most famous relation is the so-called Amati
relation \citep{Amati..2002}. This relation connects the isotropic
energy ($\Eiso$) and the peak photon energy ($\Ep$). The index of
the Amati relation ($\Ep-\Eiso$) is about $0.5$
\citep{Amati..2006,Nava..2012,Demianski..2017,Minaev..2020}.
However, it was found that this relation is not followed by some
LLGRBs. These outliers of the Amati
relation appear to follow a flatter track on the $\Ep-\Eiso$ plane
\citep{Farinelli..2021}. Previous studies showed that the off-axis
effect may play a role in this phenomenon
\citep{Ramirez..2005,Dado..2012}.

Analytical derivation of the on-axis and off-axis Amati relation
indices has been attempted in several researches
\citep{Granot..2002,Eichler..2004,Ramirez..2005,Dado..2012}.
\cite{Eichler..2004} derived the index of the Amati relation as
$0.5$ by considering a uniform, axisymmetric jet with a hole cut
out of it (i.e., a ring-shaped fireball).
\cite{Dado..2012} argued that the Amati relation should have an
index of $1/2\pm1/6$ in the framework of the so-called cannonball
model. Here, we show that the standard fireball model will also
lead to an index of $0.5$ for the on-axis Amati relation. For
the off-axis Amati relation, previous studies showed that the index
should be $1/3$ based on the effect of the viewing
angle alone \citep{Granot..2002,Ramirez..2005}. However, the effect of
the Lorentz factor was usually ignored or not fully considered for
the off-axis cases. Here we argue that the Lorentz factor is
equally responsible for determining the index of the off-axis
Amati relation. We suggest that this index should be between $1/4$
and $4/13$ after fully considering both the effect of the viewing
angle and the Lorentz factor.

Researchers have also tried to reproduce the Amati relation by
means of numerical calculations
\citep{Yamazaki..2004,Kocevski..2012,Mochkovitch..2015}. Recently,
\cite{Farinelli..2021} used an empirical comoving-frame spectrum
(a spectrum described by a smoothly broken power-law function) to
simulate the prompt emission. Their radiation flux was calculated
by averaging over the pulse duration. They found that the Amati
relation should be $\Ep\propto\Eiso^{0.5}$ for on-axis and
$\Ep\propto\Eiso^{0.25}$ for off-axis cases. Here, we consider
the detailed synchrotron spectrum instead of an empirical one in
our study, so that we can gain more insights into the detailed
physics. Furthermore, the isotropic peak luminosity $\Lp$ can be
precisely calculated in our model. As a result, in addition to
deriving the Amati relation, we can also examine the Yonetoku
relation between $\Ep$ and $\Lp$ \citep{Yonetoku..2004}, which was
previously discussed only for the on-axis cases
\citep{Zhang..2009,Ito..2019}.

Our paper is organized as follows. In Section \ref{sec:2} we
briefly describe our model. Then, in Section \ref{sec:3}, we
present an analytic derivation for the quantities of $\Ep$,
$\Eiso$, and $\Lp$ for both on-axis and off-axis cases. The
relations between these parameters are also derived. Numerical
results are presented in Section \ref{sec:4}. Next, in Section
\ref{sec:5}, Monte Carlo simulations are performed to confirm the
theoretically derived correlations. The theoretical results are
compared with the observational data in Section \ref{sec:6}.
Conclusions and discussion are presented in Section \ref{sec:7}.

\section{Jet model} \label{sec:2}

\begin{figure*}
    \centering
    \subfloat{
        \includegraphics[width=\textwidth]{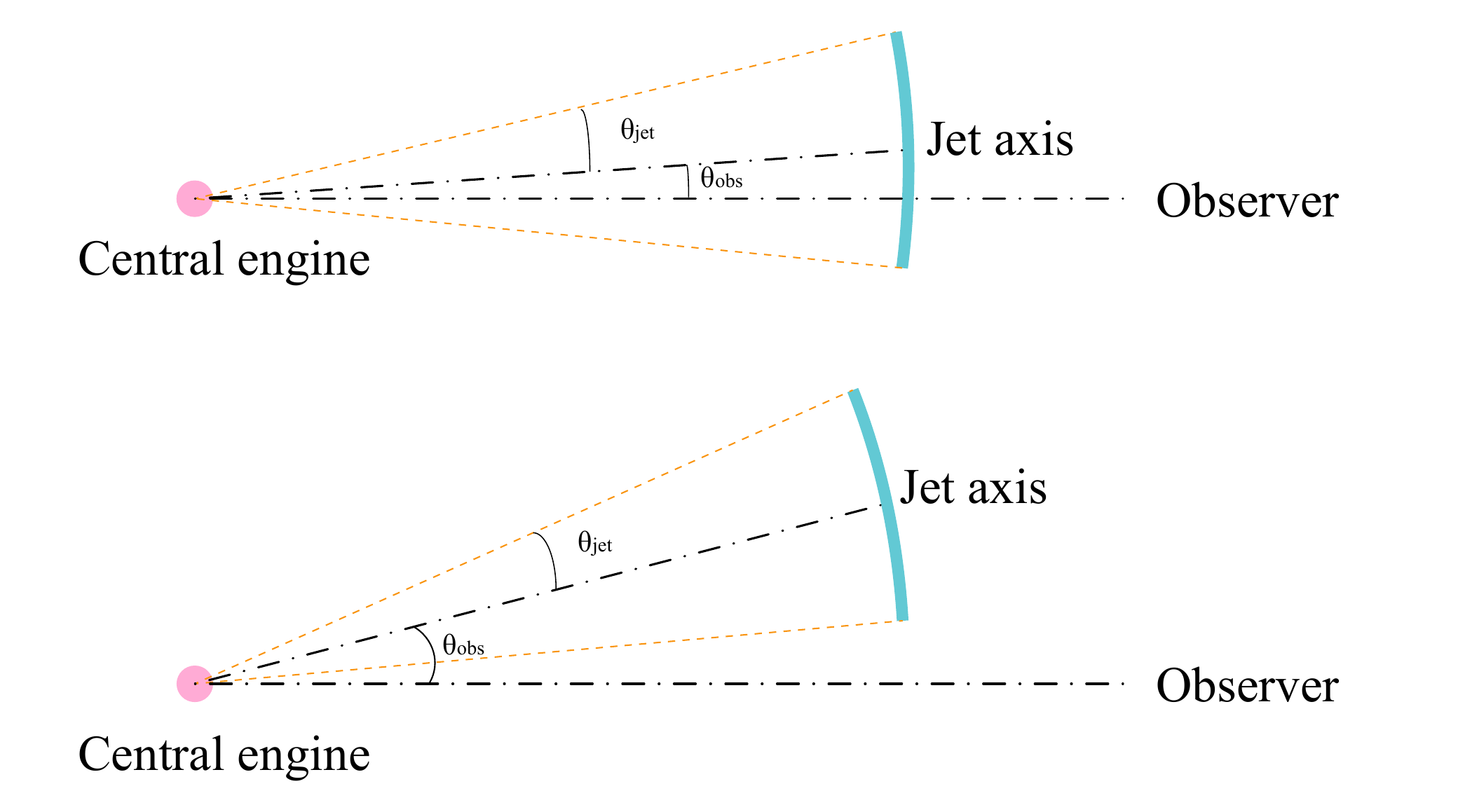}
    }

    \caption{Geometry of our jet model. The upper panel shows the case
        for an on-axis jet, where $\jet \ge \obs$. The lower panel
        shows an off-axis jet with $\jet < \obs$.} \label{fig:1}
\end{figure*}

We mainly focus on long GRBs. We used a simple top-hat jet model
similar to that of  \cite{Farinelli..2021}.
For simplicity, the radiation was assumed to only last
for a very short time interval ($\delta t$) in the local burst frame.
In other words, photons are emitted from
the shell at almost the same time. Hence the photon arrival time
is largely decided by the curvature effect. A photon
emitted at a radius $r$ with a polar angle $\theta$ will
reach the observer at
\begin{equation} \label{eq:1}
    t_{\rm{obs}}(\theta) = \frac{r}{c}(\cos \theta_{1}- \cos \theta)(1+z),
\end{equation}
where $c$ is the speed of light, and $z$ is the redshift of the
source. $\theta_{1}$ refers to the angle from which the photons first arrive. We have $\theta_{1}=0$ for an on-axis jet ($\jet \ge
\obs$), and $\theta_{1}=\obs - \jet$ for an off-axis jet ($\jet <
\obs$). The emission is produced through internal shocks, which can naturally
dissipate the kinetic energy of a baryonic fireball \citep{Zhang..2018}.
As a result, we have $r \sim \gamma^{2} d$, where $\gamma$ is the bulk Lorentz
factor of the shell, and $d$ is the initial separation between the clumps ejected
by the central engine \citep{Kobayashi..1997}. A schematic illustration of
our model is presented in Figure \ref{fig:1}.

Usually, the light curve of a single GRB pulse is composed of a fast
rising phase and an exponential decay phase \citep{Li..2021}. Complicated
processes may be involved in the prompt GRB phase, such as the hydrodynamics of the
outflow and the cooling of electrons \citep{Zhang..2007}. Analytical solutions will be too difficult to derive if these ingredients are included.
Especially in the context of the internal shock model, our assumption of
a small $\delta t$ implies that the pulse is simply produced by the collision
of two very thin shells. In realistic cases, a reverse shock may form and
propagate backward during the collision. 
The pulse profile could then be affected by the hydrodynamics of the 
flow, not only by the curvature of the emitting shell. Depending on the initial 
distribution of the Lorentz factor, the effect can remain moderate, however, and 
the two-shell model we considered is approximately valid.

Electrons accelerated by internal shocks should be in the fast-cooling regime. They will follow a distribution of \citep{Geng..2018}
\begin{equation} \label{eq:2}
    \frac{d N_{e}}{d\gamma^{\prime}_{\rm{e}}}= \left\{ \begin{array}{ll}
        C_{1}\gamma^{\prime -2}_{\rm{e}}, & \gamma^{\prime}_{\rm{m}}<\gamma^{\prime}_{\rm{e}}<\ginj, \\
        C_{2} \gamma^{\prime -p-1}_{\rm{e}}, & \ginj<\gamma^{\prime}_{\rm{e}}<\gamma^{\prime}_{\rm{max}},
    \end{array} \right.
\end{equation}
with
\begin{equation} \label{eq:3}
    \begin{array}{c}
        C_{1}=N_{\rm{e}}\left[\gamma^{\prime -1}_{\rm{m}}-\left(1-\frac{1}{p}\right)
        \gamma^{\prime -1}_{\rm{inj}}-\frac{1}{p}\left(\frac{\ginj}{\gamma^{\prime}_{\rm{max}}}\right)^{p}
        \gamma^{\prime -1}_{\rm{inj}}\right]^{-1}, \\
        C_{2}=C_{1} \gamma^{\prime -p-1}_{\rm{inj}},
    \end{array}
\end{equation}
where $\gamma^{\prime}_{\rm{e}}$ is the Lorentz factor of
electrons, and $N_{\rm{e}}$ is their total number. The spectral
index $p$ usually ranges from $2$ -- $3$ \citep{Huang..2000}.
$\gamma^{\prime}_{\rm{m}}$ and $\gamma^{\prime}_{\rm{max}}$ are
the minimum and maximum electron Lorentz factor, respectively.
$\gamma^{\prime}_{\rm{m}}$ is less than the initial Lorentz factor
of the injected electrons ($\ginj$) in the fast-cooling regime
\citep{Burgess..2020}, while $\gamma^{\prime}_{\rm{max}}$ can be
calculated approximately as $\gamma^{\prime}_{\rm{max}} \sim
10^{8} ( \frac{B^{\prime}} {1 G} )^{-0.5} $
\citep{Dai..1999,Huang..2000}, where $B^{\prime}$ is the magnetic
field strength in the comoving frame.

In the comoving frame, the synchrotron radiation power of these
electrons at frequency $\nu^{\prime}$ is \citep{Rybicki..1979}
\begin{equation} \label{eq:4}
    P^{\prime}_{\nu^{\prime}}\left(N_{\rm{e}}\right)
    = \frac{\sqrt{3} e^{3} B^{\prime}}{m_{\rm{e}} c^{2}}
    \int_{\gamma^{\prime}_{\rm{m}}}^{\gamma^{\prime}_{\rm{max}}}
    \left(\frac{d N_{\rm{e}}}{d \gamma^{\prime}_{e}}\right) F
    \left(\frac{\nu^{\prime}}{\nu_{\rm{c}}^{\prime}}\right)
    d \gamma^{\prime}_{\rm{e}},
\end{equation}
where $e$ is the electron charge, and $m_{\rm{e}}$ is the electron
mass. $\nu_{\rm{c}}^{\prime} = 3\gamma^{\prime 2}_{\rm{e}} e
B^{\prime} / 4 \pi m_{\rm{e}} c$ and $F(x)=x \int_{x}^{+\infty} K_{5/3}(k) dk$, where
$x=\frac{\nu^{\prime}}{\nu_{\rm{c}}^{\prime}}$ and $K_{5/3}(k)$ is
the Bessel function.

In the local burst frame, the angular distribution of the
radiation power is \citep{Rybicki..1979}
\begin{equation} \label{eq:5}
    \frac{d P_{\nu}}{d \Omega}=\mathcal{D}^{3}
    \frac{d P^{\prime}_{\nu^{\prime}}}{d \Omega^{\prime}}
    =\mathcal{D}^{3} \frac{P^{\prime}_{\nu^{\prime}}}{4 \pi},
\end{equation}
where $\mathcal{D}=1/[\gamma(1-\beta \cos\theta)]$ is the Doppler
factor, and $\nu^{\prime}=(1+z) \nu_{\mathrm{obs}} / \mathcal{D}$.
The radiation is assumed to be isotropic in the comoving frame.

Due to the different light-traveling time, photons emitted at
different angles ($\theta$) will be received by the observer at
different times. This is called the curvature effect. Summing up
the contribution from the whole jet, we can obtain the observed
$\gamma$-ray light curve. We first consider a time interval of
$\Delta t$ in the local burst frame. It corresponds to a thin ring
of $[\theta, \theta + \Delta \theta]$. When the electrons are
uniformly distributed in the shell, the number of electrons
in the ring is
\begin{equation} \label{eq:6}
    N_{\theta} = \frac{\phi(\theta) \sin\theta \Delta \theta}{ 2\pi
        f_{\rm{b}}}N_{\rm{tot}},
\end{equation}
where $f_{\rm{b}}=1-\cos\jet$ is the beaming factor, $N_{\rm{tot}}$ is
the total number of electrons in the shell, and $\phi(\theta)$ is
the annular angle of the ring. $N_{\rm{tot}}$ can be derived from
the isotropic-equivalent mass of the shell $m_{\rm{sh}}$ as
$N_{\rm{tot}}=2\pi f_{\rm{b}}m_{\rm{sh}}/m_{\rm{p}}$, where
$m_{\rm{p}}$ is the mass of the proton. The annular angle
$\phi(\theta)$ takes the form of
\begin{equation} \label{eq:7}
    \phi(\theta) = \left\{ \begin{array}{ll}
        2 \pi, & \theta \le | \obs-\jet |,\\
        2 \arccos (\frac{\cos \jet -\cos \obs \cos \theta}{\sin \obs \sin \theta}),
        & \theta > | \obs-\jet |.\\
    \end{array} \right.
\end{equation}

Since the radiation process only lasts for a short interval of
$\delta t$, the energy emitted into one unit solid angle is
\begin{equation} \label{eq:8}
    \frac{\Delta E_{\nu}}{d \Omega} =
    \mathcal{D}^{3} \frac{P^{\prime}_{\nu^{\prime}} \delta t}{4 \pi}.
\end{equation}
This energy corresponds to a duration of $\Delta
t=r\sin\theta\Delta\theta/c$ in the local burster frame. The
luminosity is then $L_{\nu}=\Delta E_{\nu}/\Delta t$. For an
observer at distance $D_{\rm{L}}$, the observed flux at
$t_{\rm{obs}}(\theta)$ should be $F_{\nu}(\theta)=
(1+z)L_{\nu}/4\pi D_{\rm{L}}^{2}$, that is,
\begin{equation} \label{eq:9}
    F_{\nu}(\theta) = \frac{(1+z)}{4 \pi D_{\rm{L}}^{2}} \frac{c}{r}
    \frac{\mathcal{D}^{3} P^{\prime}_{\nu^{\prime}}(N_{\theta}) \delta t}
    {\sin\theta \Delta \theta}.
\end{equation}

The total isotropic energy can be calculated as
\begin{equation} \label{eq:10}
    \Eiso= \frac{4 \pi D_{\rm{L}}^{2}}{(1+z)}\int_{0}^{t_{\rm{end}}} \int_{\nu_{1}/(1+z)}^{\nu_{2}/(1+z)}F_{\nu}(\theta) d \nu d t_{\rm{obs}}(\theta),
\end{equation}
where $t_{\rm{end}}=r(\cos \theta_{1}- \cos(\jet+\obs))(1+z)/c$ is
the end time of the pulse in the observer frame. The
isotropic peak luminosity is
\begin{equation} \label{eq:11}
    \Lp= 4 \pi D_{\rm{L}}^{2} \int_{\nu_{1}/(1+z)}^{\nu_{2}/(1+z)}F_{\nu}(\theta_{\rm{p}}) d \nu,
\end{equation}
where $\theta_{\rm{p}}$ is the angle corresponding to the peak
time of the light curve. The peak photon energy $\Ep$ can be
derived from the time-integrated spectrum. In the next section, we present an analytic derivation for the prompt $\gamma$-ray emission, paying special attention to the three parameters $\Eiso$, $\Lp$, and $\Ep$.

\section{Analytic derivation of $\Ep$, $\Eiso$, and $\Lp$} \label{sec:3}

We first consider the
on-axis cases. From Equations \ref{eq:1} and \ref{eq:10}, we can derive $\Eiso$ as
\begin{equation} \label{eq:12}
    \begin{aligned}
        \Eiso &= c \delta t
        \int_{0}^{t_{\rm{end}}} \int_{\nu_{1}/(1+z)}^{\nu_{2}/(1+z)}
        \frac{\mathcal{D}^{3} P^{\prime}_{\nu^{\prime}}(N_{\theta})}{r \sin\theta \Delta \theta} d \nu d t_{\rm{obs}}(\theta) \\
        & \propto \int_{\theta_{1}}^{\theta_{2}}
        \int_{\nu_{1}/(1+z)}^{\nu_{2}/(1+z)}
        \frac{(1+z)\frac{r}{c} \sin\theta d \theta P^{\prime}_{\nu^{\prime}}(N_{\theta})}{\gamma^{3}(1-\beta \cos\theta)^{3} r \sin\theta \Delta \theta} d \nu \\
        & \propto \frac{1}{\gamma^{3}}
        \int_{\theta_{1}}^{\theta_{2}} \frac{1}{(1-\beta \cos\theta)^{3}} \int_{\nu_{1}/(1+z)}^{\nu_{2}/(1+z)} (1+z)
        \frac{ P^{\prime}_{\nu^{\prime}}(N_{\theta})}{\Delta \theta} d \nu d \theta, \\
    \end{aligned}
\end{equation}
where $\theta_{2}=\jet+\obs$. Further combining Equation
\ref{eq:4}, we have
\begin{equation} \label{eq:13}
        \begin{aligned}
        \Eiso \propto \frac{(1+z) B^{\prime}}{\gamma^{3}} \frac{\sqrt{3} e^{3} }{m_{\rm{e}} c^{2}}
        &\int_{\theta_{1}}^{\theta_{2}} \frac{1}{(1-\beta \cos\theta)^{3}} \int_{\nu_{1}/(1+z)}^{\nu_{2}/(1+z)}
        \frac{1}{\Delta \theta} \\
        & \int_{\gamma^{\prime}_{\rm{m}}}^{\gamma^{\prime}_{\rm{max}}}\left(\frac{d N_{\theta}}{d \gamma^{\prime}_{e}}\right) F\left(\frac{\nu^{\prime}}{\nu_{\rm{c}}^{\prime}}\right) d \gamma^{\prime}_{\rm{e}} d \nu d \theta.
        \\
        \end{aligned}
\end{equation}

We note that $C_{1}$ in Equation \ref{eq:3} is
largely dependent on $\gamma^{\prime}_{\rm{m}}$ and
$\gamma^{\prime}_{\rm{inj}}$. In this study, a simple case of
$\gamma^{\prime}_{\rm{m}}=\gamma^{\prime}_{\rm{inj}}/10$ is
assumed, which naturally leads to $C_{1} \propto
\gamma^{\prime}_{\rm{inj}}$. Then with Equations \ref{eq:2},
\ref{eq:3}, and \ref{eq:6}, we can further obtain
\begin{equation} \label{eq:14}
    \begin{aligned}
        \Eiso  &\propto \frac{(1+z) B^{\prime}}{\gamma^{3}}
        \int_{\theta_{1}}^{\theta_{2}} \frac{1}{(1-\beta \cos\theta)^{3}} \int_{\nu_{1}/(1+z)}^{\nu_{2}/(1+z)}
        \frac{1}{\Delta \theta} \\ & \quad \left[\int_{\gamma^{\prime}_{\rm{m}}}^{\gamma^{\prime}_{\rm{inj}}}
        N_{\theta} \gamma^{\prime}_{\rm{inj}} \gamma^{\prime -2}_{\rm{e}}
        F\left(\frac{\nu^{\prime}}{\nu_{\rm{c}}^{\prime}}\right) d \gamma^{\prime}_{\rm{e}} + \right. \\
        & \quad \left.\int_{\gamma^{\prime}_{\rm{inj}}}^{\gamma^{\prime}_{\rm{max}}}
        N_{\theta} \gamma^{\prime p}_{\rm{inj}} \gamma^{\prime -p-1}_{\rm{e}}
        F\left(\frac{\nu^{\prime}}{\nu_{\rm{c}}^{\prime}}\right) d \gamma^{\prime}_{\rm{e}}
        \right] d \nu d \theta \\
        & \propto \frac{(1+z) B^{\prime} m_{\rm{sh}}}{\gamma^{3}}
        \int_{\theta_{1}}^{\theta_{2}}
        \frac{\phi(\theta) \sin\theta}{(1-\beta \cos\theta)^{3}} \\ & \quad \left[\int_{\nu_{1}/(1+z)}^{\nu_{2}/(1+z)}  \int_{\gamma^{\prime}_{\rm{m}}}^{\gamma^{\prime}_{\rm{inj}}}
        \gamma^{\prime}_{\rm{inj}} \gamma^{\prime -2}_{\rm{e}}
        F\left(\frac{\nu^{\prime}}{\nu_{\rm{c}}^{\prime}}\right) d \gamma^{\prime}_{\rm{e}} d \nu  + \right. \\
        & \quad \left.\int_{\nu_{1}/(1+z)}^{\nu_{2}/(1+z)} \int_{\gamma^{\prime}_{\rm{inj}}}^{\gamma^{\prime}_{\rm{max}}}
        \gamma^{\prime p}_{\rm{inj}} \gamma^{\prime -p-1}_{\rm{e}}
        F\left(\frac{\nu^{\prime}}{\nu_{\rm{c}}^{\prime}}\right) d \gamma^{\prime}_{\rm{e}} d \nu
        \right] d \theta.
    \end{aligned}
\end{equation}

For an on-axis observer, most of the observed photons are emitted
by electrons within a small angle around the line of sight.
According to Equation \ref{eq:7}, it is safe for
us to simply take $\phi(\theta)$ as $2\pi$. Changing the integral
order in Equation \ref{eq:14}, we obtain
\begin{equation} \label{eq:15}
    \begin{aligned}
        \Eiso & \propto \frac{(1+z) B^{\prime} m_{\rm{sh}}}{\gamma^{3}}
        \int_{\theta_{1}}^{\theta_{2}}
        \frac{\sin\theta}{(1-\beta \cos\theta)^{3}} \\
        & \quad \left[ \int_{\gamma^{\prime}_{\rm{m}}}^{\gamma^{\prime}_{\rm{inj}}}
        \int_{\nu_{1}/(1+z)}^{\nu_{2}/(1+z)}
        \gamma^{\prime}_{\rm{inj}} \gamma^{\prime -2}_{\rm{e}}
        F\left(\frac{\nu^{\prime}}{\nu_{\rm{c}}^{\prime}}\right) d \gamma^{\prime}_{\rm{e}} d \nu  + \right.  \\
        & \quad \left.\int_{\gamma^{\prime}_{\rm{inj}}}^{\gamma^{\prime}_{\rm{max}}}
        \int_{\nu_{1}/(1+z)}^{\nu_{2}/(1+z)}
        \gamma^{\prime p}_{\rm{inj}} \gamma^{\prime -p-1}_{\rm{e}}
        F\left(\frac{\nu^{\prime}}{\nu_{\rm{c}}^{\prime}}\right) d \gamma^{\prime}_{\rm{e}} d \nu
        \right] d \theta \\
        & \propto \frac{(1+z) B^{\prime} m_{\rm{sh}}}{\gamma^{3}}
        \int_{\theta_{1}}^{\theta_{2}}
        \frac{\sin\theta}{(1-\beta \cos\theta)^{3}} \\
        & \quad \left[ \int_{\gamma^{\prime}_{\rm{m}}}^{\gamma^{\prime}_{\rm{inj}}}
        \gamma^{\prime}_{\rm{inj}} \gamma^{\prime -2}_{\rm{e}}
        \frac{\gamma^{\prime 2}_{\rm{e}} B^{\prime}}{(1+z)\gamma(1-\beta \cos\theta)} \right. \\
        & \quad \int_{\frac{\nu^{\prime}_{1}}{\nu^{\prime}_{\rm{c}}}/(1+z)}^{\frac{\nu^{\prime}_{2}}{\nu^{\prime}_{\rm{c}}}/(1+z)}
        F\left(\frac{\nu^{\prime}}{\nu_{\rm{c}}^{\prime}}\right)
        d (\frac{\nu^{\prime}_{1}}{\nu^{\prime}_{\rm{c}}})
        d \gamma^{\prime}_{\rm{e}} + \\
        & \quad \int_{\gamma^{\prime}_{\rm{inj}}}^{\gamma^{\prime}_{\rm{max}}}
        \gamma^{\prime p}_{\rm{inj}} \gamma^{\prime -p-1}_{\rm{e}}
        \frac{\gamma^{\prime 2}_{\rm{e}} B^{\prime}}{(1+z)\gamma(1-\beta \cos\theta)} \\
        & \quad \left. \int_{\frac{\nu^{\prime}_{1}}{\nu^{\prime}_{\rm{c}}}/(1+z)}^{\frac{\nu^{\prime}_{2}}{\nu^{\prime}_{\rm{c}}}/(1+z)}
        F\left(\frac{\nu^{\prime}}{\nu_{\rm{c}}^{\prime}}\right)
        d (\frac{\nu^{\prime}_{1}}{\nu^{\prime}_{\rm{c}}})
        d \gamma^{\prime}_{\rm{e}}
        \right] d \theta \\
        & \propto \frac{B^{\prime 2} m_{\rm{sh}}}{\gamma^{4}}
        \int_{\theta_{1}}^{\theta_{2}}
        \frac{\sin\theta}{(1-\beta \cos\theta)^{4}}
        \left[ \int_{\gamma^{\prime}_{\rm{m}}}^{\gamma^{\prime}_{\rm{inj}}}
        k_{0} \gamma^{\prime}_{\rm{inj}}
        d \gamma^{\prime}_{\rm{e}} + \right. \\
        & \ \ \ \ \  \left.\int_{\gamma^{\prime}_{\rm{inj}}}^{\gamma^{\prime}_{\rm{max}}}
        k_{0} \gamma^{\prime p}_{\rm{inj}} \gamma^{\prime -p+1}_{\rm{e}}
        d \gamma^{\prime}_{\rm{e}}
        \right] d \theta. \\
    \end{aligned}
\end{equation}

In the last step above, the integral of
$F\left(\frac{\nu^{\prime}}{\nu_{\rm{c}}^{\prime}}\right)$ has
been simplified as a constant ($k_{0}$). Equation \ref{eq:15} can
be further reduced as
\begin{equation} \label{eq:16}
    \begin{aligned}
        \Eiso &\propto \frac{B^{\prime 2} m_{\rm{sh}}}{\gamma^{4}}
        \int_{\theta_{1}}^{\theta_{2}}
        \frac{\sin\theta}{(1-\beta \cos\theta)^{4}}
        \left[ k_{1} \gamma^{\prime}_{\rm{inj}}
        (\gamma^{\prime}_{\rm{inj}}-\gamma^{\prime}_{\rm{m}}) + \right. \\
        & \quad \left. k_{2} \gamma^{\prime p}_{\rm{inj}}
        (\gamma^{\prime 2-p}_{\rm{inj}}-\gamma^{\prime 2-p}_{\rm{max}})
        \right] d \theta \\
        &\propto \frac{B^{\prime 2} \gamma^{\prime 2}_{\rm{inj}} m_{\rm{sh}}}{\gamma^{4}}
        \int_{\theta_{1}}^{\theta_{2}}
        \frac{\sin\theta}{(1-\beta \cos\theta)^{4}} d \theta, \\
    \end{aligned}
\end{equation}
where $k_{1}$ and $k_{2}$ are integration constants. When $\theta
<< 1$ and $\beta \sim 1$, we have $1/(1-\beta \cos\theta) \sim
2\gamma^{2}/(\theta^{2}\gamma^{2}+1)$. The integral in Equation
\ref{eq:16} can therefore be approximated as
\begin{equation} \label{eq:17}
    \begin{aligned}
        \int_{\theta_{1}}^{\theta_{2}}
        \frac{\sin\theta}{(1-\beta \cos\theta)^{4}} d \theta
        & \approx \int_{\theta_{1}}^{\theta_{2}}
        \frac{\theta}{(\frac{\theta^{2}\gamma^{2}+1}{2\gamma^{2}})^{4}} d \theta \\
        & = 16 \gamma^{6} \int_{\theta_{1}}^{\theta_{2}}
        \frac{\theta\gamma}{(\theta^{2}\gamma^{2}+1)^{4}} d (\theta\gamma) \\
        & = 16 \gamma^{6}
        \left.-\frac{1}{6((\theta\gamma)^{2}+1)^{3}}\right|_{\theta_{1}=0} ^{\theta_{2}=\obs+\jet\gg\frac{1}{\gamma}} \\
        & \approx \frac{8}{3} \gamma^{6}. \\
    \end{aligned}
\end{equation}
Finally, combining Equations \ref{eq:16} and \ref{eq:17}, it is
easy to obtain $\Eiso \propto \gamma^{\prime 2}_{\rm{inj}}
m_{\rm{sh}} B^{\prime 2} \gamma^{2}$.

Now we derive the peak luminosity of $\Lp$. Since the main
difference between Equations \ref{eq:10} and \ref{eq:11} is the
integral of time, we can first consider the luminosity
corresponding to a particular angle $\theta$ as
\begin{equation} \label{eq:18}
    \begin{aligned}
        L(\theta) & \propto \frac{B^{\prime 2} \gamma^{\prime 2}_{\rm{inj}} m_{\rm{sh}}}{r \gamma^{4}}
        \frac{1}{(1-\beta \cos\theta)^{4}} \\
        & \propto \frac{B^{\prime 2} \gamma^{\prime 2}_{\rm{inj}} m_{\rm{sh}}}{d \gamma^{6}}
        \frac{1}{(\frac{\theta^{2}\gamma^{2}+1}{2\gamma^{2}})^{4}} \\
        & \propto \frac{B^{\prime 2} \gamma^{\prime 2}_{\rm{inj}} m_{\rm{sh}} \gamma^{2}}{d}
        \frac{1}{(\theta^{2}\gamma^{2}+1)^{4}}.
    \end{aligned}
\end{equation}
The flux peaks at $\theta_{\rm{p}} \sim 0$, thus the peak
luminosity is $\Lp=L(\theta_{\rm{p}}) \propto \gamma^{\prime
    2}_{\rm{inj}} m_{\rm{sh}} d ^{-1} B^{\prime 2} \gamma^{2}$.

The peak photon energy in the observer frame can be derived by
considering the standard synchrotron emission mechanism,
\begin{equation} \label{eq:19}
    \begin{aligned}
        E_{\rm{p, obs}} & \approx \frac{1}{(1+z)} \frac{1}{\gamma(1-\beta \cos \theta)} \frac{3 h e B^{\prime}}{4 \pi m_{e} c} \gamma^{\prime 2}_{\rm{inj}} \\
        & \propto \frac{B^{\prime} \gamma^{\prime 2}_{\rm{inj}}}{(1+z)} \frac{2 \gamma}{\theta^{2} \gamma^{2}+1} \\
        & \approx \frac{B^{\prime} \gamma^{\prime 2}_{\rm{inj}}}{(1+z)} \frac{2 \gamma}{\theta_{\rm{p}}^{2} \gamma^{2}+1} \\
        & \propto \frac{1}{(1+z)} \gamma B^{\prime} \gamma^{\prime 2}_{\rm{inj}}. \\
    \end{aligned}
\end{equation}
Therefore, the peak photon energy in the comoving frame is
$\Ep=(1+z)E_{\rm{p, obs}} \propto \gamma^{\prime 2}_{\rm{inj}}
B^{\prime} \gamma$.

Next, we consider the off-axis cases. Again, we
first focus on $\Lp$. For an off-axis jet, the main difference is
that $\phi(\theta)$ can no longer be taken as $2\pi$ . Instead,
it should be calculated as $\phi(\theta)\equiv 2 \arccos
(\frac{\cos \jet -\cos \obs \cos \theta}{\sin \obs \sin \theta})$.
Equation \ref{eq:18} now becomes
\begin{equation} \label{eq:20}
    L(\theta) \propto \frac{B^{\prime 2} \gamma^{\prime 2}_{\rm{inj}} m_{\rm{sh}}}{d \gamma^{6}} \frac{1}{(1-\beta \cos\theta)^{4}} 2 \arccos (\frac{\cos \jet -\cos \obs \cos \theta}{\sin \obs \sin \theta}).
\end{equation}
Here $\theta\in(\obs-\jet, \obs+\jet)$. We define
$R(\theta)=\frac{\cos \jet -\cos \obs \cos \theta}{\sin \obs \sin
    \theta}$. Then it is obvious that $R(\theta) \in (0,1)$, which
leads to the approximation $\arccos(R(\theta)) \sim \sqrt{2}
\sqrt{1-R(\theta)}$. We note that $R(\theta)$ can be simplified as
\begin{equation} \label{eq:21}
    \begin{aligned}
        R(\theta) & = \frac{\cos \jet -\cos \obs \cos \theta}{\sin \obs \sin \theta} \\
        & \approx \frac{1-\jet^{2}/2 - (1-\obs^{2}/2) (1-\theta^{2}/2)}{ \obs \theta} \\
        & \approx \frac{\obs^{2} - \jet^{2} + \theta^{2}}{2 \obs \theta}. \\
    \end{aligned}
\end{equation}
When $\theta>\frac{1}{\gamma}$, we have $\frac{1}{(1-\beta
    \cos\theta)^{4}} \sim
\frac{16\gamma^{8}}{(\theta^{2}\gamma^{2}+1)^{4}} \sim
\frac{16}{\theta^{8}}$. Combining Equations \ref{eq:20} and
\ref{eq:21}, we obtain
\begin{equation} \label{eq:22}
    \begin{aligned}
        L(\theta) & \propto \frac{B^{\prime 2} \gamma^{\prime 2}_{\rm{inj}} m_{\rm{sh}}}{d \gamma^{6}} \frac{1}{\theta^{8}} \sqrt{1-R(\theta)} \\
        & \propto \frac{B^{\prime 2} \gamma^{\prime 2}_{\rm{inj}} m_{\rm{sh}}}{d \gamma^{6}} \frac{1}{\theta^{8}}
        \sqrt{\frac{\jet^{2} - (\obs-\theta)^{2}}{2 \obs\theta}}. \\
    \end{aligned}
\end{equation}
In this case, $\Lp$ will peak at
$\theta_{\rm{p}}=\frac{1}{15}(16\obs-\sqrt{255\jet^{2}+\obs^{2}})
\sim \frac{16}{15}(\obs-\jet)$. We note that the condition of
$\theta_{\rm{p}} > \frac{1}{\gamma}$ leads to
$\frac{16}{15}(\obs-\jet)>\frac{1}{\gamma}$. Under this condition,
$\Lp$ can be further written as
\begin{equation} \label{eq:23}
    \begin{aligned}
        \Lp & \propto \frac{B^{\prime 2} \gamma^{\prime 2}_{\rm{inj}} m_{\rm{sh}}}{d \gamma^{6}} \frac{1}{\theta_{\rm{p}}^{8}}
        \sqrt{\frac{\jet^{2} - (\obs-\theta_{\rm{p}})^{2}}{2 \obs\theta_{\rm{p}}}} \\
        & \propto \frac{B^{\prime 2} \gamma^{\prime 2}_{\rm{inj}} m_{\rm{sh}}}{d \gamma^{6}} \frac{1}{\theta_{\rm{p}}^{8.5}}
        \sqrt{\frac{\jet^{2} - (\obs-\theta_{\rm{p}})^{2}}{2 \obs}} \\
        & \propto \frac{B^{\prime 2} \gamma^{\prime 2}_{\rm{inj}} m_{\rm{sh}}}{d \gamma^{6}} (\obs-\jet)^{-8.5} I(\obs,\jet,\theta=\theta_{\rm{p}}). \\
    \end{aligned}
\end{equation}
The last factor is
$I(\obs,\jet,\theta=\theta_{\rm{p}})=\sqrt{\frac{\jet^{2} -
        (\obs-\theta_{\rm{p}})^{2}}{2 \obs}}$. Its effect is not important
compared with $(\obs-\jet)^{-8.5}$, and therefore, we ignore this factor for
simplicity. Finally, we obtain $\Lp \propto \gamma^{\prime
    2}_{\rm{inj}} m_{\rm{sh}} d ^{-1} B^{\prime 2} \gamma^{-6}
(\obs-\jet)^{-8.5}$.

Now we continue to calculate $\Eiso$. Since most of the
$\gamma$-ray energy is released in the decay stage of the light
curve, we only need to integrate the energy over a $\theta$ ranging
from $\theta_{\rm{p}}$ to $\obs+\jet$, that is,
\begin{equation} \label{eq:24}
    \begin{aligned}
        \Eiso & \propto \frac{B^{\prime 2} \gamma^{\prime 2}_{\rm{inj}} m_{\rm{sh}}}{\gamma^{4}}
        \int_{\theta_{\rm{p}}}^{\obs+\jet}
        \frac{\sin\theta}{(1-\beta \cos\theta)^{4}} \\
        & \quad 2 \arccos (\frac{\cos \jet -\cos \obs \cos \theta}{\sin \obs \sin \theta}) d \theta \\
        & \propto \frac{B^{\prime 2} \gamma^{\prime 2}_{\rm{inj}} m_{\rm{sh}}}{\gamma^{4}} \int_{\theta_{\rm{p}}}^{\obs+\jet}
        \frac{1}{\theta^{7}} \sqrt{\frac{\jet^{2} - (\obs-\theta)^{2}}{2 \obs\theta}} d \theta \\
        & \propto \frac{B^{\prime 2} \gamma^{\prime 2}_{\rm{inj}} m_{\rm{sh}}}{\gamma^{4}} \int_{\theta_{\rm{p}}}^{\obs+\jet}
        \frac{1}{\theta^{7.5}} I(\obs,\jet,\theta) d \theta \\
        & \approx \frac{B^{\prime 2} \gamma^{\prime 2}_{\rm{inj}} m_{\rm{sh}}}{\gamma^{4}} \int_{\theta_{\rm{p}}}^{\obs+\jet}
        \frac{1}{\theta^{7.5}} I(\obs,\jet,\theta=\theta_{p}) d \theta \\
        & \propto \frac{B^{\prime 2} \gamma^{\prime 2}_{\rm{inj}} m_{\rm{sh}}}{\gamma^{4}} I(\obs,\jet,\theta=\theta_{p})
        \left.\theta^{-6.5}\right|_{\theta_{\rm{p}}}^{\obs+\jet}  \\
        & \propto \frac{B^{\prime 2} \gamma^{\prime 2}_{\rm{inj}} m_{\rm{sh}}}{\gamma^{4}} \theta_{\rm{p}}^{-6.5}. \\
    \end{aligned}
\end{equation}
Here the value of $I(\obs,\jet,\theta)$ will not change
significantly in the range from $\theta_{\rm{p}}$ to $\obs+\jet$.
Hence it can be taken as a constant for simplicity. Since
$\theta_{\rm{p}} \sim \frac{16}{15}(\obs-\jet)$, we obtain the result
as $\Eiso \propto \gamma^{\prime 2}_{\rm{inj}} m_{\rm{sh}}
B^{\prime 2} \gamma^{-4} (\obs-\jet)^{-6.5}$.

Finally, we calculate $\Ep$. Similar to Equation \ref{eq:19},
the peak energy in the rest frame is
\begin{equation} \label{eq:25}
    \begin{aligned}
        \Ep & \approx B^{\prime} \gamma^{\prime 2}_{\rm{inj}}
        \frac{2 \gamma}{\theta_{\rm{p}}^{2} \gamma^{2}+1} \\
        & \approx B^{\prime} \gamma^{\prime 2}_{\rm{inj}}
        \frac{2 \gamma}{(\obs-\jet)^{2} \gamma^{2}} \\
        & \propto \frac{B^{\prime} \gamma^{\prime 2}_{\rm{inj}}}{\gamma}(\obs-\jet)^{-2}. \\
    \end{aligned}
\end{equation}
We see that $\Ep \propto \gamma^{\prime 2}_{\rm{inj}} B^{\prime}
\gamma^{-1} (\obs-\jet)^{-2}$.

To summarize, in the on-axis cases, we have
\begin{equation} \label{eq:26}
    \begin{array}{ll}
        \Ep & \propto \gamma^{\prime 2}_{\rm{inj}} B^{\prime} \gamma, \\
        \Eiso & \propto \gamma^{\prime 2}_{\rm{inj}} m_{\rm{sh}} B^{\prime 2} \gamma^{2}, \\
        \Lp & \propto \gamma^{\prime 2}_{\rm{inj}} m_{\rm{sh}} d ^{-1} B^{\prime 2} \gamma^{2},
    \end{array}
\end{equation}
while in the off-axis cases ($\obs-\jet>1/\gamma$), we obtain
\begin{equation} \label{eq:27}
    \begin{array}{ll}
        \Ep & \propto \gamma^{\prime 2}_{\rm{inj}} B^{\prime} \gamma^{-1} (\obs-\jet)^{-2}, \\
        \Eiso & \propto \gamma^{\prime 2}_{\rm{inj}} m_{\rm{sh}} B^{\prime 2} \gamma^{-4} (\obs-\jet)^{-6.5}, \\
        \Lp & \propto \gamma^{\prime 2}_{\rm{inj}} m_{\rm{sh}} d ^{-1} B^{\prime 2} \gamma^{-6} (\obs-\jet)^{-8.5}.
    \end{array}
\end{equation}

The equations derived above are very simple but intriguing. There are
eight input parameters in our model: $\gamma$, $B^{\prime}$,
$\ginj$, $\jet$, $\obs$, $m_{\rm{sh}}c^{2}$, $d$, and the electron
spectral index $p$. Seven of them are closely related to the prompt
emission parameters of $\Ep$, $\Eiso$, and $\Lp$, while $p$ is largely
irrelevant. Of all the input parameters, $\gamma$ plays the most
important role in affecting the observed spectral peak energies and
fluxes for on-axis cases \citep{Ghirlanda..2018}. Usually, $\gamma$ may
vary in a wide range for different GRBs, from lower than 100 to more
than 1000. In contrast, $\gamma^{\prime}_{\rm{inj}}$ and
$m_{\rm{sh}}$ only vary in much narrower ranges. Taking
$\gamma^{\prime}_{\rm{inj}}$ and $m_{\rm{sh}}$ approximately
as constants, from Equation \ref{eq:26}, we can
easily derive the on-axis Amati relation as $\Ep \propto \Eiso^{0.5}$
and the on-axis Yonetoku relation as $\Ep \propto \Lp^{0.5}$. In both
relations, the power-law indices are 0.5, which is largely consistent
with observations \citep{Nava..2012,Demianski..2017}.

In the off-axis cases, previous studies mainly focused on the
effect of the variation of the viewing angle among different GRBs.
Consequently, the power-law index of the off-axis Amati relation
is derived as $1/3$ based on the sharp-edge homogeneous jet
geometry \citep{Granot..2002,Ramirez..2005}. Here, from our
Equation~\ref{eq:27}, we see that $\Eiso$ and $\Lp$ sensitively
depend on both the viewing angle and the Lorentz factor. This
indicates that both $\gamma$ and $\obs$ are important parameters
that will affect the slope of the Amati relation and the Yonetoku
relation. In Equation~\ref{eq:27}, if only the variation
of the viewing angle is considered (i.e., the
indices of $(\obs-\jet)$ in Equation \ref{eq:27} are taken into account, but the
item of $\gamma$ is omitted), an index of $4/13$ will be derived for the
off-axis Amati relation, which is very close to the previous result of
$1/3$. When the combined effect of varying $\gamma$ and
$\obs$ is included, the off-axis Amati relation is derived as
$\Ep \propto \Eiso^{1/4 \sim 4/13}$, and the corresponding
off-axis Yonetoku relation is $\Ep \propto \Lp^{1/6 \sim
4/17}$. The power-law indices become smaller than in the on-axis cases.

\section{Numerical results} \label{sec:4}

Some approximations have been made in deriving Equations
\ref{eq:26} and \ref{eq:27}. In this section, we carry out
numerical simulations to confirm whether the above analytical
derivations are correct. For convenience, a set of standard
values were taken for the eight input parameters in our
simulations, as shown in Table \ref{tab:1}. When studying the
effect of one particular parameter, we only changed this
parameter, but fixed all other parameters at the standard values.

\begin{table}[h!]
    \renewcommand{\thetable}{\arabic{table}}
    \centering
    \caption{Standard values assumed for the eight input parameters} \label{tab:1}
    \begin{tabular}{cccccccc}
        \hline
        \hline
        $\gamma$ & $B^{\prime}$ & $\ginj$ & $\jet$ & $\obs$ & $p$ & $m_{\rm{sh}}c^{2}$ & $d$ \\
        & (G) &   & (rad) & (rad) \tablefootmark{a} &  & (erg) \tablefootmark{b} & (cm) \\
        \hline
        100 & 10 & 1e5 & 0.1 & 0.0(0.2) & 2.5 & 1e52 & 1e10  \\
        \hline
    \end{tabular}
    \tablefoot{
        \tablefoottext{a}{We set the standard value as $\obs=0.0$ for an on-axis jet and $\obs=0.2$ for an off-axis jet.}
        \tablefoottext{b}{The value of $m_{\rm{sh}}$ here is the isotropic equivalent mass.}
    }
\end{table}

To examine the correctness of our analytical derivations, we chose
one input parameter as a variable, and standard values were taken
for all other input parameters so that the dependence of $\Ep$,
$\Eiso$, and $\Lp$ on that particular parameter can be illustrated.

\begin{figure*}
    \centering
    \includegraphics[width=\textwidth]{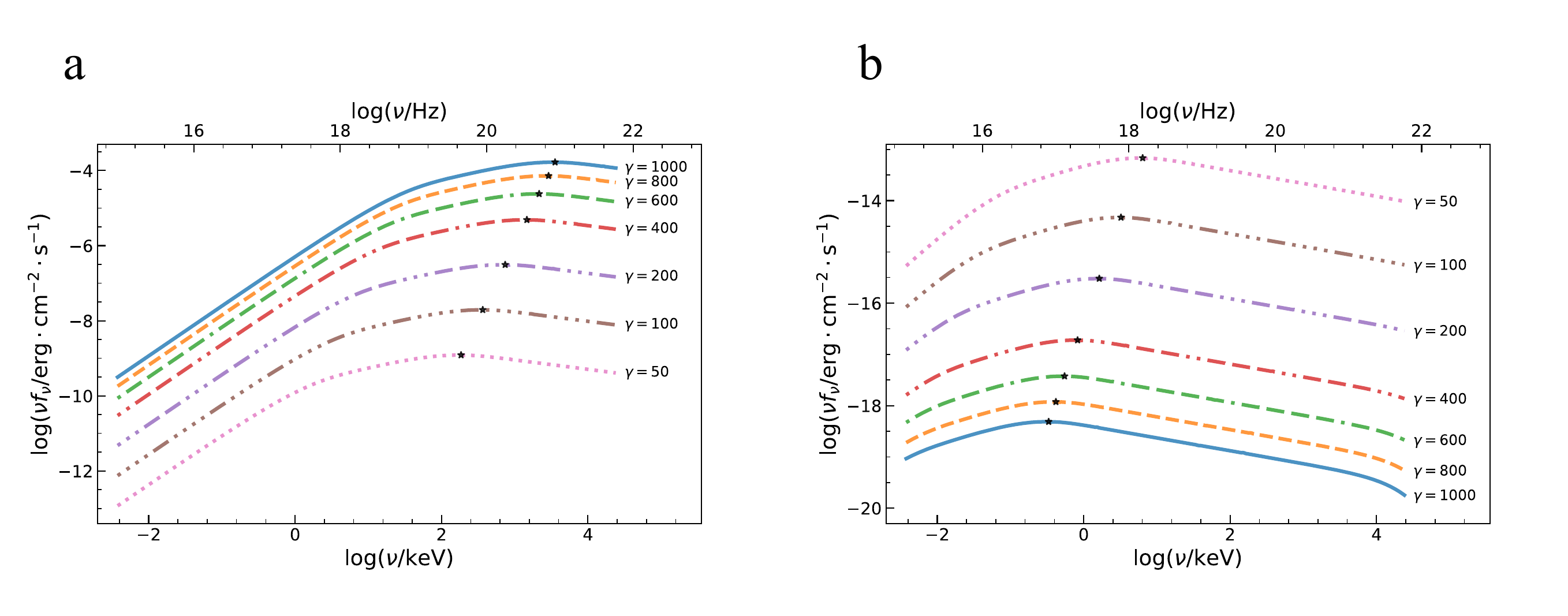}
    \caption{Time-integrated spectrum for different $\gamma$ values
        (marked in the figure). Panel {\bf a} Spectrum of on-axis cases.
 Panel        {\bf b} Spectrum of off-axis cases. The black star on each curve marks
        the position of the peak energy ($\Ep$). } \label{fig:2}
\end{figure*}

We first consider the effect of the bulk Lorentz factor $\gamma$
for both on-axis and off-axis cases. Here we let $\gamma$ vary
between 50 --- 1000. Figure \ref{fig:2} shows the time-integrated
spectrum obtained for different $\gamma$. The peak photon energy
is correspondingly marked with a black star on the curve.

For on-axis cases ($\jet\ge\obs$), the numerical results are
plotted in the left panel of Figure \ref{fig:2}. When the value of
$\gamma$ increases, the peak photon energy and the flux increase
simultaneously. However, a different trend is found for the
off-axis cases ($\jet<\obs$), as shown in the right panel of
Figure \ref{fig:2}. Both the flux and $\Ep$ decrease as $\gamma$
increases. The flux is significantly lower than that of
the on-axis cases. The results are consistent with those of
\cite{Farinelli..2021}.

\begin{figure*}
    \centering
    \includegraphics[width=\textwidth]{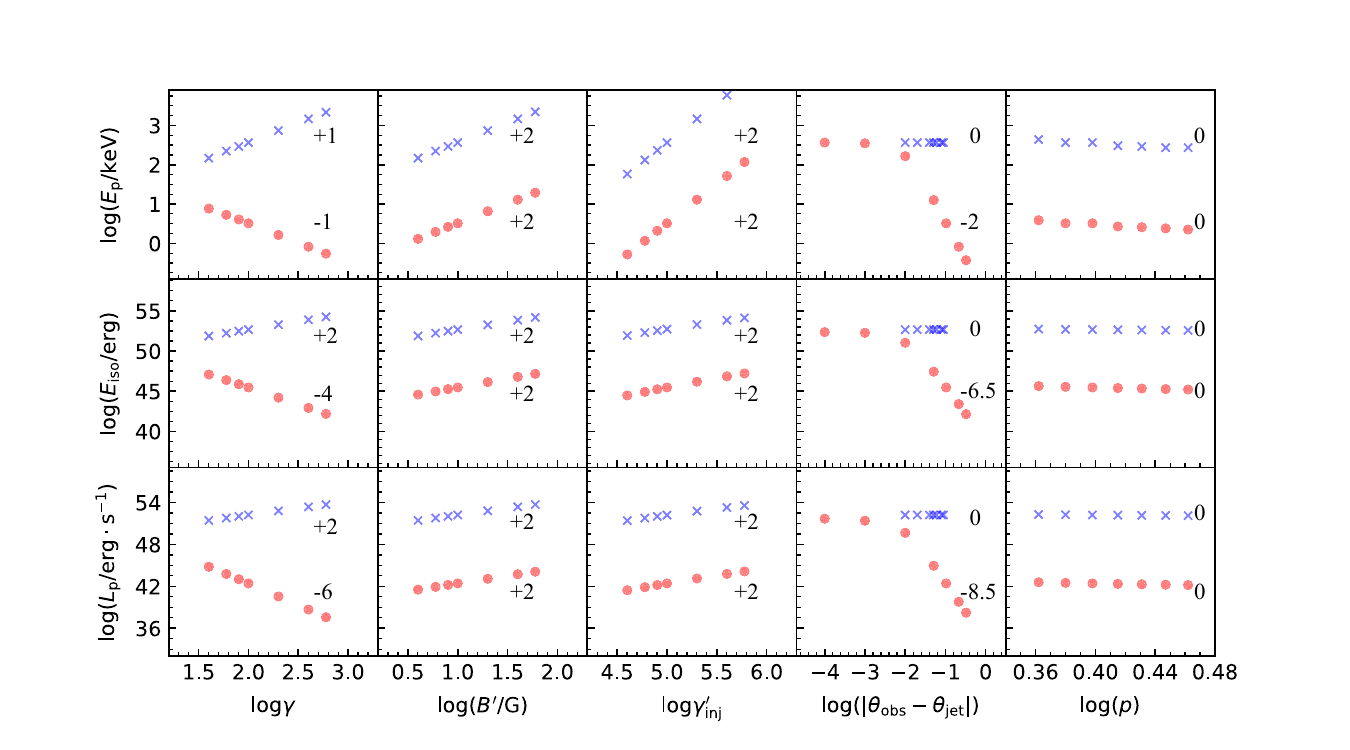}
    \caption{ Numerical results showing the dependence of
            the prompt emission features on the input parameters.
        The calculated prompt emission parameters
        ($\Ep$, $\Eiso$, and $\Lp$) are plotted against the
        input parameters ($\gamma$, $B^{\prime}$, $\ginj$, $|\obs-\jet|$,
        and $p$) for both the on-axis cases (with cross symbols)
        and the off-axis cases (with circle symbols).
        For comparison, the power-law indices derived from our
        analytical solutions are marked correspondingly in each panel. } \label{fig:3}
\end{figure*}

Figure \ref{fig:2} clearly shows that both $\Ep$ and the flux are
positively correlated with $\gamma$ in the on-axis cases, but they
are negatively correlated with $\gamma$ in the off-axis cases.
Next, we wish to determine the precise relation between the prompt
emission parameters and the input parameters. Hence, we plot
$\Ep$, $\Eiso$, and $\Lp$ as functions of $\gamma$, $B^{\prime}$,
$\ginj$, $|\obs-\jet|$, and $p$ in Figure \ref{fig:3}.

Generally, the simulation results are well consistent
with our analytical results of Equations \ref{eq:26} and
\ref{eq:27}. Especially $\Ep$, $\Eiso$, and $\Lp$ have a positive
dependence on the Lorentz factor $\gamma$ in the on-axis cases,
but they have a negative dependence on $\gamma$ in the off-axis
cases. It is also interesting
to note that the prompt emission parameters are nearly independent
of $\obs-\jet$ in the on-axis cases, since the line of sight is
always within the jet cone. However, in the off-axis cases, the
dependence of the prompt emission parameters on $\obs-\jet$ can be
described as a two-phase behavior. When $\obs-\jet\ll1/\gamma$,
the prompt emission parameters are independent of $\obs-\jet$. When $\obs-\jet>1/\gamma$, however, they decrease sharply with the increase
of $\obs-\jet$, as indicated in Equation \ref{eq:27}. Finally,
the last column of Figure \ref{fig:3} shows that the parameter
$p$ has little impact on the prompt emission parameters.

\section{Monte Carlo simulations} \label{sec:5}

In this section, we perform Monte Carlo simulations to generate a large
number of mock GRBs to further test the existence of the Amati relation and
the Yonetoku relation. For this purpose, we performed Monte Carlo simulations
as follows. First, a group of eight input parameters is generated randomly, assuming that each parameter follows a particular distribution. This group of parameters defines a mock GRB. Second, our model is applied to numerically calculate the corresponding values of $\Ep$, $\Eiso$, and $\Lp$ for this mock GRB. The above two steps are repeated until we obtain a large sample of mock GRBs. Finally, the mock bursts are plotted on the $\Ep-\Eiso$ plane and the $\Ep-\Lp$ plane, and a best-fit correlation is obtained for them.

For convenience, we designate the distribution function of a
particular parameter $a$ as $D(a)$. Here $a$ refers to the input
parameters (i.e., $\gamma$, $B^{\prime}$, $\ginj$, $\jet$, $\obs$,
$p$, $m_{\rm{sh}}c^{2}$, and $d$). Three kinds of distributions
are adopted in this study: (i) A normal Gaussian distribution,
which is noted as $N(\mu,\sigma)$, where $\mu$ and $\sigma$ refer
to the mean value and standard deviation, respectively. (ii) A
uniform distribution, which is noted as $U(\rm{min},\rm{max})$,
where $\rm{min}$ and $\rm{max}$ define the range of the parameter. Finally,
(iii) a log-normal distribution, noted as $D(\log(a))$, which
means that $\log (a)$ follows a normal Gaussian distribution of
$N(\mu,\sigma)$. The detailed distribution for each of the input
parameters adopted in our Monte Carlo simulations is described
below.

The ranges of some input parameters, such as $\gamma$, $\jet$, and
$\obs$, can be inferred from observations. For the bulk Lorentz
factor, a distribution of $D(\log(\gamma))=N(2.2,0.8)$ was assumed,
which is consistent with observational
constraints \citep{Liang..2010,Ghirlanda..2018}. As for the angle
parameters, we set $\jet$ as a uniform distribution of
$D(\jet)=U(0.04,0.2)$ for the on- and off-axis
cases \citep{Frail..2001,Wang..2018}. The distribution of $\obs$ was
taken as $D(\obs)=U(0.0,0.2)$ in the on-axis cases, together with
the restriction of $\jet\ge\obs$. On the other hand, in the
off-axis cases, $\obs$ was taken as $D(\obs)=U(0.1,0.5)$, together
with $\jet<\obs$.

Other parameters cannot be directly measured from observations.
Their distributions are then taken based on some theoretical
assumptions. The isotropic ejecta mass is usually thought to
range from $10^{-3}M_{\odot}$ to $10^{-1}M_{\odot}$ for a typical
long GRB. Hence we assumed that $m_{\rm{sh}}c^{2}$ has a log-normal
distribution. We set its mean value as $10^{-2}M_{\odot}c^{2}$ and
its $2\sigma$ range as $10^{-3}M_{\odot}c^{2}$ to
$10^{-1}M_{\odot}c^{2}$, that is, $D(\log(m_{\rm{sh}}c^{2} /
\rm{erg})) = N(52.2,0.5)$. The distribution of $d$ was taken as
$D(\log(d/\rm{cm})) = N(10,0.2)$, which, combined with the above
distribution of $\gamma$, gives the range of the internal shock
radius as $r \sim \gamma^{2} d \sim 10^{13}-10^{16}$ cm. It is
largely consistent with the theoretical expectation of $10^{11}-10^{17}$ cm \citep{Zhang..2018}.

Some parameters ($B^{\prime}$, $\ginj$, $p$) concern the
microphysics of relativistic shocks. The comoving magnetic field
strength ($B^{\prime}$) is hard to estimate from
observations \citep{Zhang..2011,Burgess..2020}. Here, a log-normal
distribution with a mean value of 10 G and a standard deviation of
0.4 G was assumed for it, that is, $D(\log(B^{\prime} / \rm{G})) =
N(1,0.4)$. The mean value of $\ginj$ was taken as $10^{5}$, with a
distribution of $D(\log(\ginj)) = N(5,0.2)$ \citep{Burgess..2020}.
For the parameter $p$, we simply assumed a normal distribution of
$D(p)=N(2.5,0.4)$.

With the above distribution functions of the input parameters, two
samples of mock GRBs were generated through Monte Carlo
simulations. One sample included 200 on-axis events, and the other
sample included 200 off-axis cases. $\Ep$, $\Eiso$, and $\Lp$ were
calculated for each mock event.

\begin{figure*}
    \centering
    \includegraphics[width=\textwidth]{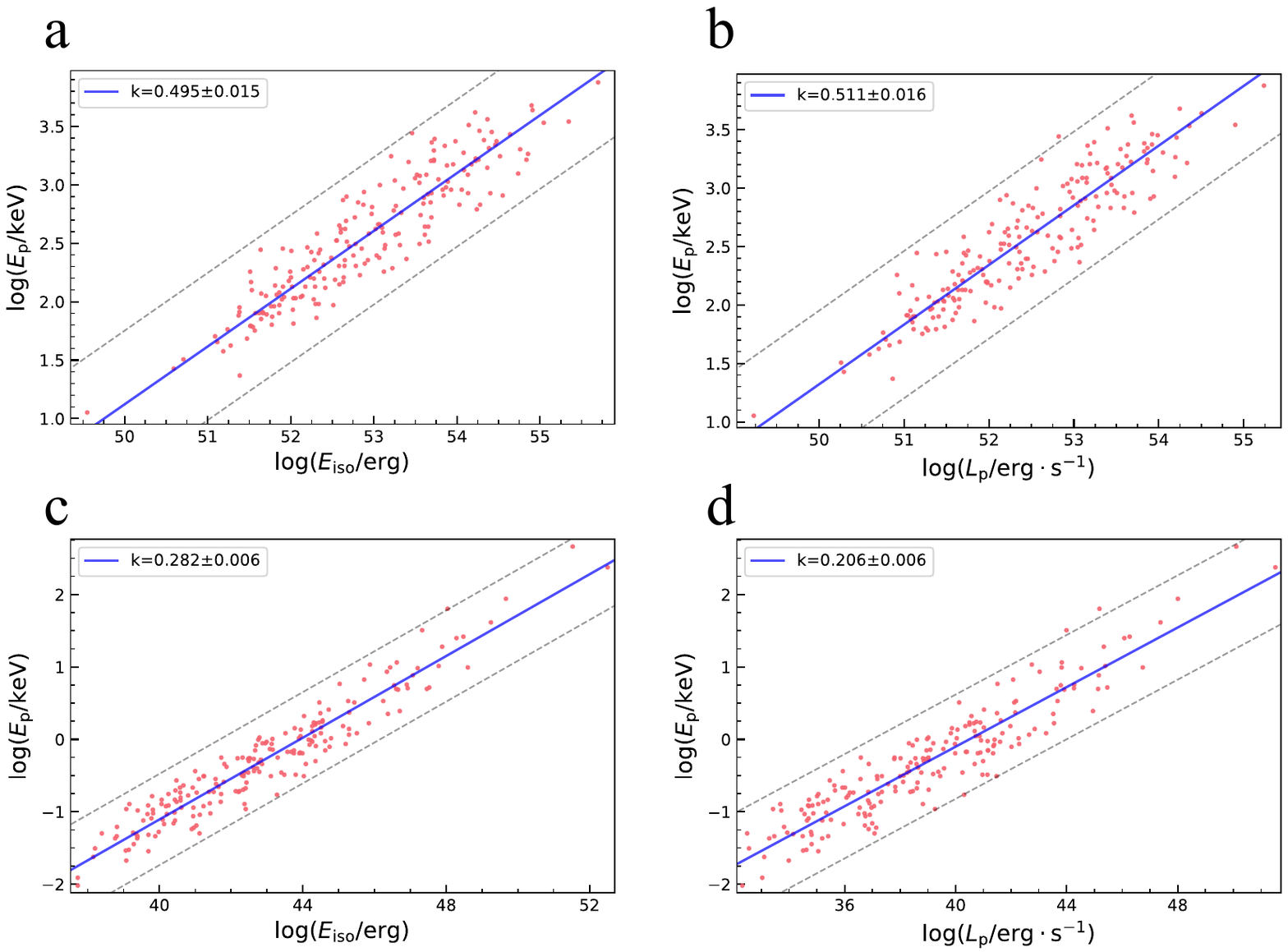}
    \caption{Monte Carlo simulation results. {\bf (a,b),} Distribution of simulated bursts on the $\Ep-\Eiso$
        plane and the $\Ep-\Lp$ plane for the on-axis cases.
        {\bf (c,d),} Distribution of simulated bursts on the
        $\Ep-\Eiso$ plane and the $\Ep-\Lp$ plane for off-axis cases.
        The solid lines show the best-fitting results for the
        simulated data points, and the dashed lines represent
        the $3\sigma$ confidence level. To conclude, for on-axis
        bursts, the best-fit Amati relation is $\Ep \propto
        \Eiso^{0.495\pm0.015}$ and the best-fit Yonetoku relation
        is $\Ep \propto \Lp^{0.511\pm0.016}$. For off-axis bursts,
        the best-fit Amati relation is $\Ep \propto
        \Eiso^{0.282\pm0.006}$, while the best-fit Yonetoku
        relation is $\Ep \propto \Lp^{0.206\pm0.006}$. } \label{fig:4}
\end{figure*}

The distribution of the simulated bursts on the
$\Ep-\Eiso$ and $\Ep-\Lp$ planes is shown in Figure \ref{fig:4}.
In the on-axis cases, the best-fit slope is
$0.495\pm0.015$ for the $\Ep-\Eiso$ correlation, and it is
$0.511\pm0.016$ for the $\Ep-\Lp$ correlation. These two
indices are very close to our theoretical results of $0.5$,
proving credibility for our analytical derivations. The
intrinsic scatters of both relations are about 0.21, indicating
that the correlations are rather tight.

\begin{figure*}
    \centering
    \includegraphics[width=\textwidth]{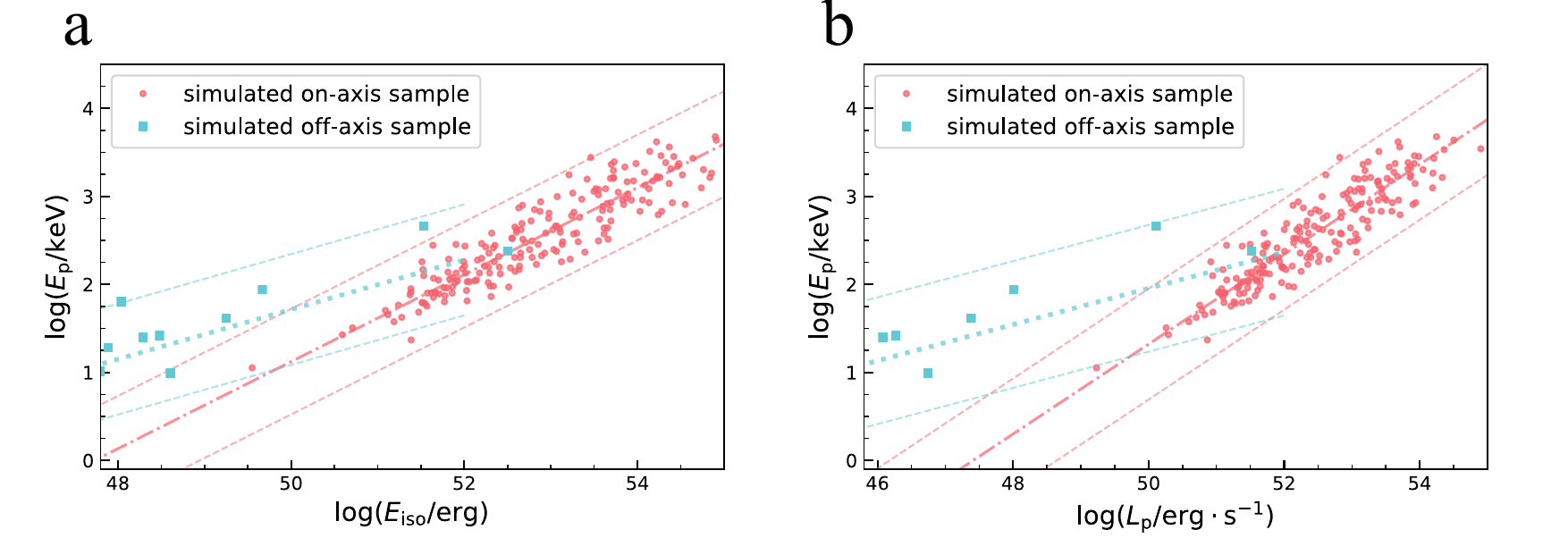}
    \caption{Distribution of simulated off-axis bursts as
            compared with that of on-axis ones. {\bf a,} All mock bursts on
        the $\Ep-\Eiso$ plane. {\bf b,} All mock bursts on
        the $\Ep-\Lp$ plane. The dash-dotted line and the dotted
        line correspond to the best-fit result for on- and
        off-axis samples, respectively. The dashed lines represent
        the corresponding $3\sigma$ confidence level. Most of the off-axis bursts are too weak to appear
        in the two panels here. } \label{fig:5}
\end{figure*}

In the off-axis cases, the best-fit slope is $0.282 \pm 0.006$ for
the $\Ep-\Eiso$ correlation, which is consistent with our expected
range of $1/4 \sim 4/13$ for the off-axis Amati relation.
Similarly, on the $\Ep-\Lp$ plane, the best-fit slope is
$0.206\pm0.006$, which also agrees well with our theoretical range
of $1/6 \sim 4/17$ for the off-axis Yonetoku relation. Another
interesting point in Figure \ref{fig:4}c and \ref{fig:4}d is that
most of the simulated off-axis GRBs have an $\Eiso$ lower than
$10^{48}$ erg and an $\Lp$ lower than $10^{46}$ erg s$^{-1}$.
Only about 5\% of our simulated off-axis GRBs have greater $\Eiso$
and $\Lp$. We thus argue that off-axis GRBs are usually
LLGRBs. It also indicates that the majority
of currently observed normal GRBs should be on-axis GRBs. In
Figure \ref{fig:5}, all the simulated on-axis and off-axis bursts
are plotted in the same panel. Only
a very small number of off-axis events are strong enough to
be detected. Their distribution obviously deviates from that of
the on-axis events on both the $\Ep-\Eiso$ and $\Ep-\Lp$ plane.

\begin{figure*}
    \centering
    \includegraphics[width=\textwidth]{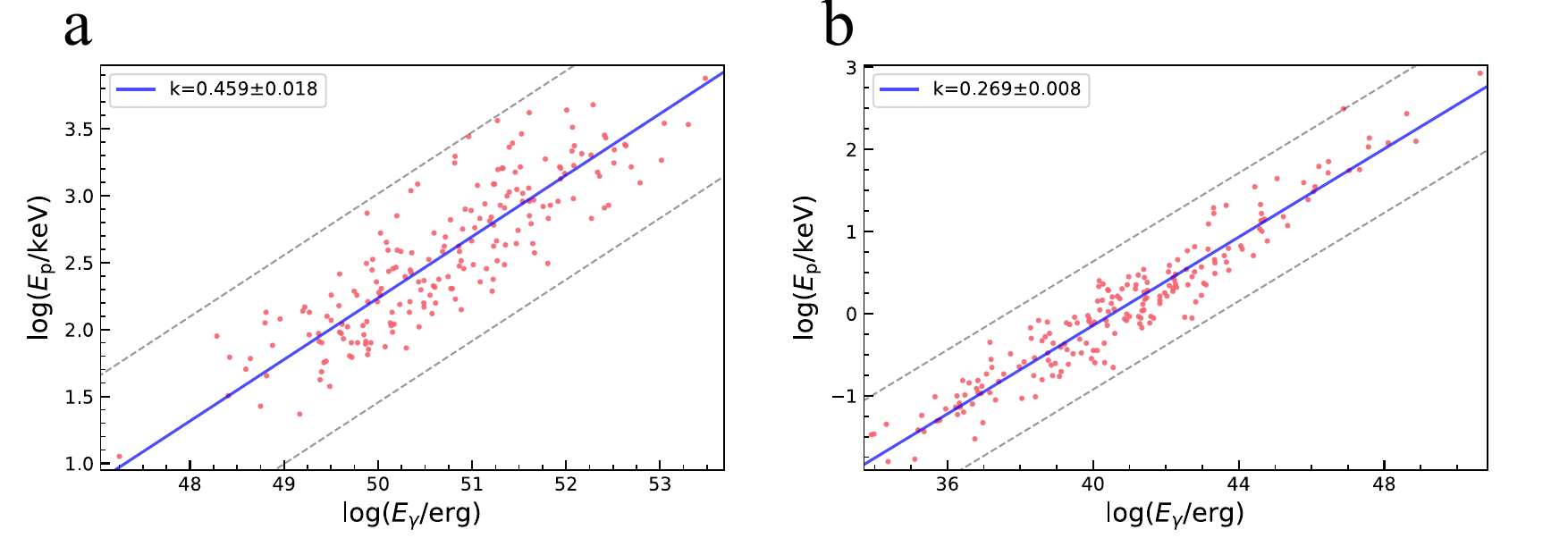}
    \caption{Distribution of the mock GRBs on the $\Ep-E_{\gamma}$
        plane for the on-axis case. Panel {\bf a} On-axis case.
Panel        {\bf b} Off-axis case. The solid line shows the best-fit result for the simulated data points, and the dashed lines represent the $3\sigma$ confidence
        level. The on-axis Ghirlanda relation is $\Ep \propto E_{\gamma}^{0.459 \pm 0.018}$, and the off-axis Ghirlanda relation reads $\Ep \propto E_{\gamma}^{0.269 \pm 0.008}$.} \label{fig:6}
\end{figure*}

Using the mock bursts, we can also test the Ghirlanda
relation \citep{Ghirlanda..2004}, which links $\Ep$ with the
collimation-corrected energy of $E_{\gamma}=(1-\cos\jet)\Eiso$.
The distribution of the simulated bursts on the $\Ep  -
E_{\gamma}$ plane is shown in Figure \ref{fig:6}. The data points are best fit by a power-law function of $\Ep
\propto E_{\gamma}^{0.459 \pm 0.018}$ for the on-axis case (Figure \ref{fig:6}a), with an intrinsic scatter
of $0.263 \pm 0.031$. This agrees well with the recently
updated Ghirlanda relation of $\Ep \propto
E_{\gamma}^{0.44\pm0.07}$ for a sample of 55 observed
GRBs \citep{Wang..2018}. For the off-axis case (Figure \ref{fig:6}b), we find $\Ep
\propto E_{\gamma}^{0.269 \pm 0.008}$, with a larger scatter of $0.272 \pm 0.029$. This indicates that the Ghirlanda relation may be
largely connected with the Amati relation, but compared with the
latter, the slope of the Ghirlanda relation is slightly flatter,
and the data points are also obviously more scattered. It is difficult task to compare the off-axis Ghirlanda relation with observations because it needs the precise determination of both the beaming angle and the viewing angle for a number of off-axis GRBs, which themselves are rather dim.

\section{Comparison with observations} \label{sec:6}

In Section \ref{sec:5} we illustrated the simulated
$\Ep-\Eiso$ and $\Ep-\Lp$ relations for both on-axis and off-axis
cases. We now proceed to compare our results with
observational data.

A sample containing 172 long GRBs was used for this purpose, 162 of which were collected from the previous study by \cite{Demianski..2017} and \cite{Nava..2012}. $\Ep$
and $\Eiso$ parameters are available for all these 162 events.
The $\Lp$ parameter is available for only 45 GRBs \citep{Nava..2012}. Additionally, we collected another
10 bursts that were argued to be possible outliers of the
normal Amati relation in previous studies. It has been suggested
that they might be off-axis
GRBs \citep{Yamazaki..2003b,Ramirez..2005}. The observational data
of these 10 events are listed in Table \ref{tab:2}.

\begin{figure*}
    \centering
    \includegraphics[width=\textwidth]{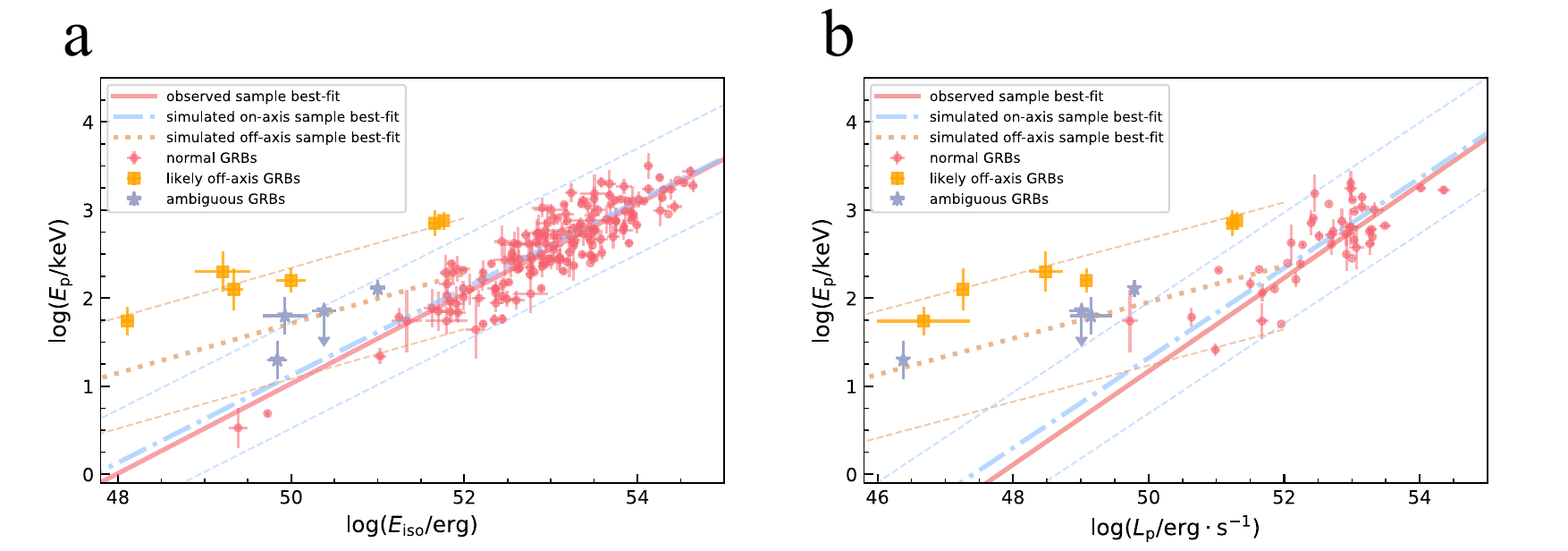}
    \caption{ Distribution of all the observed GRBs. Panel {\bf a}
        $\Ep$ vs. $\Eiso$. Panel {\bf b} $\Ep$ vs. $\Lp$. The dash-dotted
        lines represent our theoretical results for on-axis bursts,
        and the thick dotted line shows our theoretical results for
        off-axis bursts. The $3\sigma$ confidence levels are marked
        correspondingly. The solid lines are the best-fit results for
        normal GRBs. } \label{fig:7}
\end{figure*}

All the observed GRB samples are plotted on the $\Ep-\Eiso$ and
$\Ep-\Lp$ planes in Figure \ref{fig:7}. Most of the GRBs are well
consistent with our on-axis results, thus they both follow the
Amati relation and the Yonetoku relation.

On the $\Ep-\Eiso$ plane (Figure \ref{fig:7}a), four of the ten possible outliers are located between
our on- and off-axis lines. It
is thus difficult to judge whether these events are on-axis
or off-axis GRBs simply from the Amati relation. However, the
remaining six bursts can only be matched by our off-axis line.
They should indeed be off-axis events. Figure
\ref{fig:4}c and \ref{fig:4}d clearly show that most
off-axis GRBs should be LLGRBs with $\Eiso<10^{48}$ erg, with only
a small portion falling in the range of $10^{48}-10^{52}$ erg (see
also Figure \ref{fig:5}a). This is easy to understand. The input
parameters should take some extreme values for an off-axis GRB to
be strong enough. Especially these off-axis events should
generally have a low $\obs-\jet$ value (see Equation
\ref{eq:27}), which means that they are still slightly off-axis. In Figure \ref{fig:7}a, only a small number of GRBs are
outliers of the on-axis Amati relation. This low event rate is
also consistent with our simulation results.


For the $\Ep-\Lp$ correlation (Figure \ref{fig:7}b), we note
that almost all the ten possible outliers clearly deviate from the
on-axis Yonetoku relation. From combining the Amati relation
and the Yonetoku relation, we therefore argue that the ambiguous sample
should also be off-axis GRBs. Furthermore, Figure \ref{fig:7}
seems to indicate that the Yonetoku relation may be a better tool
for distinguishing between on- and off-axis GRBs.

\section{Conclusions and discussion} \label{sec:7}

The prompt emission of GRBs was studied for both on- and
off-axis cases. Especially the three prompt emission parameters $\Ep$, $\Eiso$, and $\Lp$ were considered. Their dependence
on the input model parameters was obtained via both analytical
derivations and numerical simulations, which are consistent with
each other. We confirmed that $\Ep$, $\Eiso$, and $\Lp$ are
independent of $\obs$ and $\jet$ as long as our line of sight is
within the homogeneous jet cone (i.e., the on-axis cases).
However, they strongly depend on the value of $\obs-\jet$ when the
jet is viewed off-axis. Additionally, their dependence on $\gamma$
is very different for on- and off-axis cases, as shown
in Equations \ref{eq:26} and \ref{eq:27}. Through Monte Carlo
simulations, we found that the Amati relation is $\Ep \propto
\Eiso^{0.495\pm0.015}$ in on-axis cases. Correspondingly, the
Yonetoku relation is $\Ep \propto \Lp^{0.511\pm0.016}$. On the
other hand, in off-axis cases, the Amati relation is $\Ep \propto
\Eiso^{0.282\pm0.006}$ and the Yonetoku relation is $\Ep \propto
\Lp^{0.206\pm0.006}$. The simulated samples were directly compared
with the observational samples. We found that they are well
consistent with each other in the slopes and intrinsic scatters.
The $\Ep-E_{\gamma}$ relation was also tested and was found to be $\Ep \propto E_{\gamma}^{0.459\pm0.018}$ for on-axis GRBs and
$\Ep \propto E_{\gamma}^{0.269\pm0.008}$ for off-axis GRBs.

We have focused on the empirical correlations of long GRBs in the above
analysis. Here we present some discussion of short GRBs.
Generally speaking, long GRBs usually contain many pulses
(tens or even up to hundreds) in their light curves, while short
GRBs contain much fewer pulses. For short GRBs, we can therefore perform a similar analysis, but only
when the number of pulses considered is reduced significantly. In our
model, the total energy release ($\Eiso$) is proportional to the
number of pulses, while the peak luminosity ($\Lp$) and the
spectral parameter $\Ep$ are nearly irrelevant. In other
words, a short GRB will have a much smaller $\Eiso$, but its $\Lp$
and $\Ep$ are largely unchanged. As a result, short GRBs will
follow a parallel track on the $\Ep-\Eiso$ plane as compared with
long GRBs, the only difference is that their $\Eiso$ are
significantly smaller. Alternatively speaking, the power-law index
of the Amati relation should be the same for short GRBs and long
GRBs, consistent with currently available observations
\citep{Zhang..2009,ZhangZB..2018}. As for the Yonetoku relation on
the $\Ep-\Lp$ plane, we argue that it should be almost identical
for both long and short GRBs. However, the number of currently
available short GRBs that is suitable for exploring the Yonetoku relation
(i.e., for which $\Ep$ and $\Lp$ are measured) in detail is still too
small.

A simple top-hat jet geometry was adopted in this study. It should
be noted that the GRB outflow might be a structured
jet \citep{Ioka..2018}. A recent study suggested that $\Ep$
and the flux will not change much, regardless of the choice of a
top-hat jet or a not-too-complicated structured
jet \citep{Farinelli..2021}. However, in some more complicated
scenarios, structured jets may further have an angle-dependent
energy density \citep{Lamb..2021}. They may be choked jets or even
in jet-cocoon systems \citep{Ioka..2018,Mooley..2018,Troja..2019}.
The empirical correlations of GRBs in these situations are beyond
the scope of this study and could be further considered in the
future.

In our calculations, the prompt emission was assumed to be due to
nonthermal radiation from internal shocks. Other
mechanisms such as photosphere radiation and magnetic reconnection above
the photosphere may also contribute to the observed flux of GRBs.
Generally, there are two types of photosphere models: nondissipative 
photosphere models, and dissipative photosphere models. The former can  
produce a narrow quasi-Planckian component and could account for the thermal 
component observed in some GRBs \citep{Peer..2008,Beloborodov..2011}, although 
this component does not play a dominant role in most GRBs \citep{Guiriec..2011,Axelsson..2012}.
In the dissipative photosphere models, various subphotosphere dissipation 
processes \citep{Rees..2005}, especially Comptonization, may affect the 
spectrum \citep{Thompson..1994,Peer..2006,Peer..2011,Veres..2012}, which may lead to a nonthermal 
spectrum. Some authors have even used this model to explain the Band spectrum of 
GRBs \citep{Thompson..1994,Beloborodov..2013,Lundman..2013}. Reconnection above the
photosphere has also been considered, especially in the view of the ICMART model,
focusing on Poynting-flux-dominated outflows \citep{Zhang..2011,Zhang..2014}. 
It is still unclear how commonly photosphere radiation and magnetic 
reconnection are involved in the prompt phase of GRBs. If these radiation 
mechanisms are included, the derivation will become much more complicated, 
but it deserves a trial in the future. 

The Amati and Yonetoku relations are essential for us to understand
the nature of GRBs. They can also help us probe the high-redshift
universe \citep{Xu..2021,Hu..2021,Zhao..2022,Jia..2022,Deng..2023}. 
Our study shows that they
are due to the ultra-relativistic effect of highly collimated
jets. To be more specific, they mainly result from the
variation in the Lorentz factor in different events. It may
provide useful insights for a better understanding of these
empirical correlations.

\begin{acknowledgements}
We would like to thank the anonymous referee for helpful suggestions.
This study is supported by the National Natural Science
Foundation of China (Grant Nos. 12233002, 12041306, 12147103, U1938201,
U2031118, 12273113, 11903019, 11833003), by the National Key R\&D Program of China
(2021YFA0718500), by National SKA Program of China No.
2020SKA0120300, and by the Youth Innovations and Talents Project of Shandong Provincial Colleges and Universities (Grant No. 201909118).
\end{acknowledgements}

\nocite{*}
\bibliographystyle{aa}
\bibliography{bibtex}

\begin{table*}[h!]
        \renewcommand{\thetable}{\arabic{table}}
        \centering
        \caption{Some possible outliers} \label{tab:2}
        \begin{tabular}{cccccc}
                \hline
                \hline
                Likely off-axis sample&&&&& \\
                \hline
                GRB Name & z & $\log(\Ep)$ \tablefootmark{a} & $\log(\Eiso)$ \tablefootmark{b} & $\log(\Lp)$ \tablefootmark{b} & Refs \tablefootmark{c} \\
                (Units) &   & (keV) & ($10^{52}$ erg) & ($10^{51}$ erg s$^{-1}$) &  \\
                \hline
                GRB 980425 & 0.0085 & 1.74$\pm$0.16 & -3.89$\pm$0.07 & -4.32$\pm$0.69 & 2, 7 \\
                GRB 031203 & 0.106 & 2.2$\pm$0.14 & -2$\pm$0.17 & -1.92$\pm$0.09 & 1, 2 \\
                GRB 061021 & 0.346 & 2.85$\pm$0.15 & -0.34$\pm$0.08 & 0.24$\pm$0.11 & 4, 11 \\
                GRB 080517 & 0.09  & 2.3$\pm$0.23 & -2.79$\pm$0.32 & -2.52$\pm$0.25 & 3, 5 \\
                GRB 140606B & 0.384 & 2.88$\pm$0.1 & -0.24$\pm$0.02 & 0.3$\pm$0.02 & 10 \\
                GRB 171205 & 0.0368 & 2.1$\pm$0.24 & -2.66$\pm$0.11 & -3.74$\pm$0.09 & 10, 12 \\
                \hline
                Ambiguous sample&&&&& \\
                \hline
                GRB Name & z & $\log(\Ep)$ & $\log(\Eiso)$ & $\log(\Lp)$ & Refs \\
                (Units) &   & (keV) & ($10^{52}$ erg) & ($10^{51}$ erg s$^{-1}$) &  \\
                \hline
                GRB 100316D & 0.059 & 1.3$\pm$0.22 & -2.16$\pm$0.11 & -4.62$\pm$0.01 & 7, This work \\
                GRB 120422A & 0.28253 & $<$1.86 & -1.62$\pm$0.14 & -1.99$\pm$0.18 & 8, 6 \\
                GRB 150818 & 0.282 & 2.11$\pm$0.1 & -1$\pm$0.09 & -1.21$\pm$0.06 & 8, This work \\
                GRB 161219B & 0.1475 & 1.8$\pm$0.21 & -2.07$\pm$0.26 & -1.85$\pm$0.31 & 9 \\
                \hline
        \end{tabular}
        \tablefoot{
                \tablefoottext{a}{Converted to the rest frame.}
                \tablefoottext{b}{Estimated in the $1-10^{4}$ keV energy range.}
                \tablefoottext{c}{References for $\Ep$, $\Eiso$, and $\Lp$: (1) \cite{Amati..2006}; (2) \cite{Zhang..2009}; (3) \cite{Butler..2010}; (4) \cite{Nava..2012}; (5) \cite{Sun..2015}; (6) \cite{Deng..2016}; (7) \cite{Dereli..2017}; (8) \cite{Cano..2017a}; (9) \cite{Cano..2017b}; (10) \cite{Xue..2019}; (11) \cite{Minaev..2020}; (12) \cite{Dainotti..2020}. }
        }
\end{table*}

\end{document}